\documentclass[12pt]{article}
\usepackage{geometry}
\usepackage[utf8]{inputenc}
\usepackage{enumitem,amssymb}
\newlist{thematic}{itemize}{8}
\setlist[thematic]{label=$\square$}
\usepackage{graphicx}
\usepackage{pifont}

\usepackage{cmbright}
\usepackage{natbib}

\begin{document}
\noindent{\large NASA Decadal Astrobiology Research and Exploration Strategy White Paper} \\

\noindent{\bf \large Exozodiacal dust as a limitation to exoplanet imaging and spectroscopy}
\vspace{0.3cm}

\noindent Principal Author: Miles H. Currie (NASA Goddard Space Flight Center) \\
\vspace{-0.4cm}

\noindent Co-authors: John Debes, Yasuhiro Hasegawa, Isabel Rebollido, Virginie Faramaz, Steve Ertel, William Danchi, Bertrand Mennesson, Mark Wyatt, and NASA SAG23 Members \\

\noindent{\bf \large Summary} \\
In addition to planets and other small bodies, stellar systems will likely also host exozodiacal dust, or exozodi. This warm dust primarily resides in or near the habitable zone of a star, and scatters stellar light in visible to NIR wavelengths, possibly acting as a spatially inhomogeneous fog that can impede our ability to detect and characterize Earth-like exoplanets. By improving our knowledge of exozodi in the near term with strategic precursor observations and model development, we may be able to mitigate these effects to support a future search for signs of habitability and life with a direct imaging mission. This white paper introduces exozodi, summarizes its impact on directly imaging Earth-like exoplanets, and outlines several key knowledge gaps and near-term solutions to maximize the science return of future observations.

\hspace{0.5cm}

\noindent{\bf \large Introduction to exozodiacal dust} \\
Exozodiacal dust, or exozodi, is a component of stellar systems in addition to exoplanets and other small bodies. Exozodi is broadly defined as a population of warm ($\sim 300$K) and/or hot ($\sim 1000$K) dust in or near the habitable zone (HZ) of a star \citep{kralExozodiacalCloudsHot2017}. It is not necessarily spatially smooth, and can manifest as mean-motion resonance (MMR) structure with planetary bodies in the form of clumps, gaps, or warps. Exozodi emission is analogous to the zodiacal light we observe in our solar system within a few au from the Sun, which is composed of carbonaceous and/or silicate grains 1--100$\mu$m in size \citep{Grun1985}. Hot dust is thought to be dominated by sub-mircon grains, and is closer to the star \citep{Absil2006, Ertel2014}; see S. Ertel et al. (2025, forthcoming) for a review of hot dust. 

The origin of HZ dust is not well understood, and multiple mechanisms are likely to contribute to the bulk dust population. Warm dust can be created by the slow inward migration via Poynting Robertson drag of objects originating in cold outer regions analogous to our Kuiper Belt \citep{Rigley2020}, by eccentric comets evaporating near periastron \citep{beustBetaPictorisCircumstellar1990}, by a separate population of warm planetesimals analogous to our asteroid belt \citep{Rigley2020}, or by a recent catastrophic event \citep{weinbergerABSENCECOLDDUST2011}. The origin of hot dust is even less well constrained--- deposit by star-grazing comets and some trapping mechanisms are the most likely scenario, but still pose major challenges \citep{Faramaz2017, Pearce2022}. 

Observational studies have sought to constrain the density and spatial distribution of exozodi around nearby Sun-like stars \citep{beichmanNewDebrisDisks2006, defrereNullingInterferometryImpact2010, defrereDirectImagingExoEarths2012, robergeExozodiacalDustProblem2012}, and the HOSTS survey at $\sim 11 \mu$m \citep{ertelHOSTSSurveyExozodiacal2020} produced the best constraints to date: nearby Sun-like stars have a best-fit median level of 3 zodis (where 1 zodi is the density of solar system zodiacal dust), although more critically 9 and 27 zodis are the 1- and 2-$\sigma$ upper limits, respectively. Understanding exozodi in nearby systems will be critical for informing target selection for Earth-like exoplanet detection and characterization, and we will likely have to contend with exozodi contamination in images and spectra of exoEarths to some degree. Below we summarize these potential impacts before posing viable near-term recommendations to improve our knowledge of exozodi and prepare for the next generation direct imaging mission, the Habitable Worlds Observatory (HWO).

\vspace{0.5cm}

\noindent{\bf \large The impact of exozodi on exoEarth imaging and spectroscopy} \\
Exozodiacal dust is expected to be the largest source of astrophysical noise for future direct imaging observations. To first order, the exoEarth yield will be impacted even if the exozodi can be subtracted down to the Poisson noise limit. For a given planetary system, increasing the exozodi density from 3 (the HOSTS median value) to 27 (the 95\% upper limit) causes a 30--40\% loss in mission yield of exoEarth candidates for 4--8 m telescopes \citep{starkExoEarthYieldLandscape2019}. 


\vspace{0.2cm}

\noindent{\bf Detection (imaging):}
To probe whether a stellar system includes a habitable zone terrestrial planet, the first visit to the system by a direct imaging mission will likely be a broadband imaging observation. Exozodiacal dust will scatter stellar light such that the planet is embedded in a veil of extra noise that needs to be subtracted from the image. In the ideal case of a spatially smooth, face-on disk, up to $\sim1000$ zodis could be subtracted down to the Poisson limit \citep{kammererSimulatingReflectedLight2022}. However, disk inclination \citep{kammererSimulatingReflectedLight2022} and clumpy structures from MMR \citep{currieMitigatingWorstcaseExozodiacal2023} are likely to reduce this to 10--50 zodis for a majority of potential target systems.


Furthermore, unsubtracted clumpy MMR structure, which we observe in our own solar system zodiacal dust \citep{kelsallCOBEDiffuseInfrared1998}, can be a source of confusion for direct imaging. Exozodiacal dust clumps can manifest as false positive point sources when extracting planetary signals and can look very similar to a planetary PSF, especially since these clumps will likely co-orbit with the planet, trailing slightly behind. \citet{defrereDirectImagingExoEarths2012} highlights examples of planets injected into systems with different exozodi levels including MMR structure--- when extracting point source signals in cases with $>20$ zodis of clumpy dust, exozodi clumps can have higher SNR relative to the planet, and can be potentially mistaken for a planet if we have no prior knowledge of the system. Additionally, the degree to which a trailing clump can be spatially resolved, or to what degree the planet--exozodi PSFs will blend together, will depend on many factors including the PSF of the instrument, distance to the system, inclination of the system, and planetary orbital phase. 
Discriminating an exozodi clump from a planet may require spectroscopic observations, and even a detection of an MMR clump instead of a planet could still provide evidence for the presence of a fainter planetary companion. 

For inclined systems, the exozodi will forward scatter stellar light in areas where the dust is in the foreground relative to the star. Observations of planets passing through these regions will have relatively higher exozodi noise. Depending on how extreme the system inclination and exozodi density are, this could overwhelm the planet signal entirely. Fortunately, the regions of enhanced brightness due to forward scattering in inclined systems is also where the planet will be dim along its orbital path (i.e. crescent phase), and the planet will be brightest where the forward scattering is lower. Prioritizing planetary observations between gibbous phase and quadrature will minimize the forward scattering component of the exozodi noise. 

The presence of hot ($\sim1000$ K) exozodi close to the star can alter the coronagraphic speckle profile by introducing extra leakage in addition to the stellar component (S. Ertel et al., 2025, PASP, forthcoming). 
However, hot exozodi does not appear to correlate with stellar type and other populations of dust in systems, therefore knowing whether it is present before observing may be challenging. 

\begin{figure}
    \centering
    \includegraphics[width=0.8\linewidth]{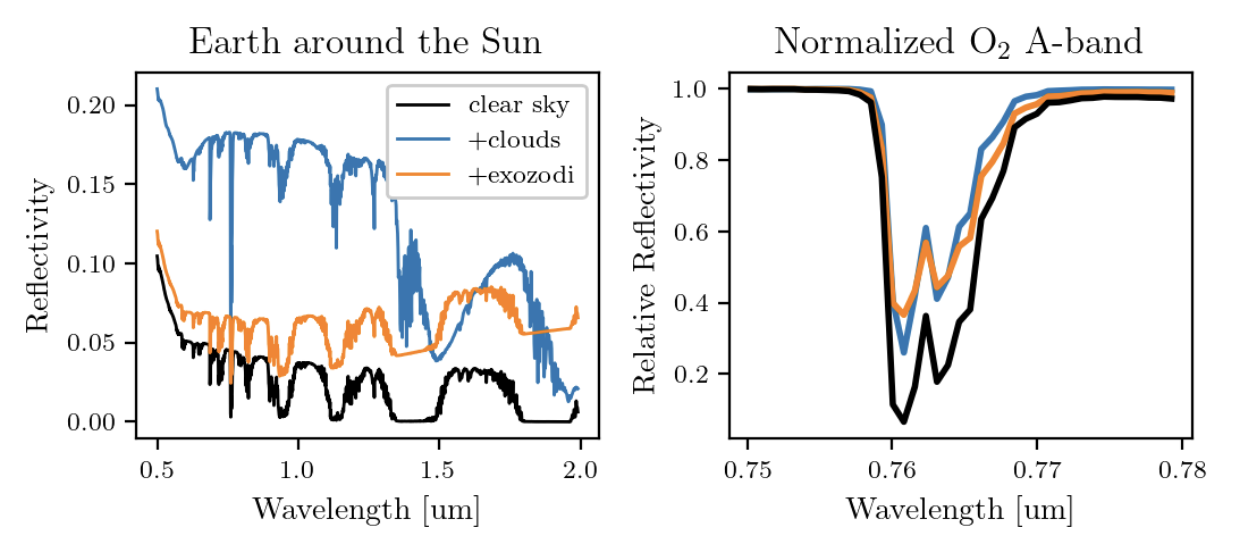}
    \caption{{\it Left:} Clear-sky Earth reflectance spectrum (black), and the addition of clouds (blue) and exozodi (orange). {\it Right:} Normalized O$_2$ A-band (biosignature) for each case. \textbf{Both exozodi and clouds can reduce the relative depth of molecular absorption features, potentially biasing inferred atmospheric parameters.} Image credit: Currie et al. (in prep)
}
    \label{fig:o2_fig}
\end{figure}

\vspace{0.2cm}

\noindent{\bf Characterization (spectroscopy):}
To first order, exozodiacal dust is a gray scatterer, but may have slight color depending on its composition--- exozodi  will introduce an additive broad, smoothly varying continuum to the observed planetary reflectance spectrum. If this extra continuum can be subtracted from the extracted planet spectrum to the Poisson noise limit, then each wavelength bin will carry an extra noise term in this best case scenario. Even so, this will affect the spectrum most where the planet is faintest (i.e. in absorption bands), and it can bias atmospheric retrieval analyses which help determine the molecular gas abundances to gain planetary environmental context. However, in the potentially more likely scenario that exozodi cannot be fully subtracted from extracted spectra, there will be some residual exozodiacal light left over that will add to the continuum. This is a problem for two reasons: 1) it could reduce the perceived relative depth of molecular bands, and 2) clouds raise the continuum in a similar way, so discriminating between these effects is challenging. 

Inferring atmospheric properties via spectroscopy may be difficult for systems with high inclinations and/or high exozodi densities. To illustrate this, Figure~\ref{fig:o2_fig} shows a simple model of a clear-sky Earth reflectance spectrum, and the effect of adding clouds and exozodi to the spectrum. The depth of the O$_2$ band relative to the continuum is reduced for the cloud/exozodi cases compared to the clear-sky case, and the relative effect is nearly the same for adding either clouds or exozodi to the clear-sky spectrum.
The level of continuum added also depends on the degree of exozodi forward scattering, and second-order effects may arise from wavelength-dependent scattering--- exozodi will not impact all spectral features uniformly (Currie et al., in prep.).

\newpage

\noindent{\bf \Large Recommendations to improve our understanding of exozodi:}\\
Exozodiacal dust will likely impact many aspects of exoEarth detection and characterization, and here we suggest several key research directions to prioritize in the near-term that will improve our understanding of exozodi in systems that may host habitable zone planets.  \\ 
\noindent{\bf 1) Investigate the population and scattering properties of warm HZ exozodi.}
Current inner working angle (IWA) limitations in visible wavelengths prevent us from imaging warm exozodi in the habitable zone. HOSTS \citep{ertelHOSTSSurveyExozodiacal2020} surveyed 23 nearby Sun-like stars---approximately eight of these are top priority targets on the ExEP HWO target list---but the survey is currently incomplete, and 30 more targets have yet to be observed pending funding. Continuing the survey would improve constraints on the median exozodi level approximately three-fold \citep{Ertel2020SPIE11446E..07E}. Additionally, the Roman Space Telescope will be equipped with a coronagraph instrument well suited for detecting exozodiacal dust at the optical wavelengths considered for HWO, and even in the case of non-detection it is expected to place upper limits complementary to LBTI for individual stars \citep{mennesson2019interplanetary, mennesson2019potential, Douglas2022}. Understanding the occurrence rates as well as the scattering properties with multi-band photometry will be critical for a future exoEarth characterization mission.

\noindent{\bf 2) If hot exozodi is present, understand how it affects coronagraphic leakage.}
A near-term observational solution to improve our understanding of hot exozodi is to use MATISSE and the upcoming NOTT instruments on VLTI to observe hot dust. MATISSE can improve constraints on dust geometry and composition, and NOTT will be capable of observing hot dust 10--50x fainter than current capabilities due to the enhanced high-contrast performance of nulling interferometry. 
Gaining more information on hot dust and its connection to warm dust will allow us to understand potential coronagraphic leakage effects, and improve our models of exozodiacal dust for post-processing removal strategies. 

\noindent{\bf 3) Constrain the origin of warm/hot exozodi.}
The origin for warm/hot exozodi is not well understood, thus exozodi models are currently limited. The composition of typical exozodiacal dust is not well known, and consequently the extent to which exozodi impacts spectral extraction is not well understood. Developing predictable models for the properties of observed exozodi, considering various delivery mechanisms, could help reveal relationships between the properties of exozodi and system architectures. These models could be used in end-to-end frameworks coupling different telescope and instrument designs with detailed treatments of dust scattering properties. By prioritizing these modeling frameworks, and including sophisticated grain models based on experimental measurements into radiative transfer simulations, we can begin to better characterize the observed properties of exozodi in visible wavelengths.

\noindent{\bf 4) Improve post-processing techniques in simulated exoEarth observations.}
Removing exozodi contamination from simulated exoEarth images to the Poisson noise limit may be possible for low-to-moderately inclined systems; however, systems that are highly inclined or edge-on may require more sophisticated techniques \citep{kammererSimulatingReflectedLight2022, currieMitigatingWorstcaseExozodiacal2023}.
If target systems have random orientations relative to us, this means that statistically up to $\sim50\%$ of targets will have inclinations requiring these better methods. We therefore recommend the development of novel post-processing techniques to handle these geometries. Possible avenues include exploring disk self-subtraction, exozodi color, or using spectral or planetary phase information to separate the exozodi from the planet signal.

\vspace{-0.5cm}

\bibliographystyle{aasjournal}
\setlength{\bibsep}{-1pt}
{
\bibliography{references}}

\end{document}